# Intraday Trading Algorithm for Predicting Cryptocurrency Price Movements Using Twitter Big Data Analysis

Vahidin Jeleskovic[1] and Stephen Mackay[2]

## Abstract

Cryptocurrencies have emerged as a novel financial asset garnering significant attention in recent years. A defining characteristic of these digital currencies is their pronounced short-term market volatility, primarily influenced by widespread sentiment polarization, particularly on social media platforms such as Twitter. Recent research has underscored the correlation between sentiment expressed in various networks and the price dynamics of cryptocurrencies. This study delves into the 15-minute impact of informative tweets disseminated through foundation channels on trader behavior, with a focus on potential outcomes related to sentiment polarization. The primary objective is to identify factors that can predict positive price movements and potentially be leveraged through a trading algorithm. To accomplish this objective, we conduct a conditional examination of return and excess return rates within the 15 minutes following tweet publication. The empirical findings reveal statistically significant increases in return rates, particularly within the initial three minutes following tweet publication. Notably, adverse effects resulting from the messages were not observed. Surprisingly, sentiments were found to have no discernible impact on cryptocurrency price movements. Our analysis further identifies that investors are primarily influenced by the quality of tweet content, as reflected in the choice of words and tweet volume. While the basic trading algorithm presented in this study does yield some benefits within the 15-minute timeframe, these benefits are not statistically significant. Nevertheless, it serves as a foundational framework for potential enhancements and further investigations.

---

[1] Humboldt Universität zu Berlin. Email: vahidin.jeleskovic@hu-berlin.de
[2] Independent researcher

# 1   Introduction

For decades inventors tried to develop a virtual currency that functions without banks as intermediaries. In 2008, the pseudonym Satoshi Nakamoto established the Bitcoin, which marked the beginning of the cryptocurrency era. Nakamoto (2008) introduced a decentralized peer-to-peer version of electronic cash that eliminated the double-spending problem[3]. However, for most economic actors in recent time cryptocurrencies are seen as a financial asset rather than a currency in the classical sense. Thus, the value of Bitcoin rapidly increased and raised the attention of governments, economists and traders. In his Ethereum White Paper, Vitalik Buterin describes Bitcoin as a "value without any backing, intrinsic value or central issuer", emphasizing that there is no real value underlying Bitcoin and altcoins (alternative cryptocurrencies).[4] Therefore, the cryptocurrencies value is connected to a proposed project, which is highly driven by speculation for future growth and reflects the investors' belief in the project.

Based on the efficient market hypothesis by Eugene F. Fama (1970), the investors' belief is affected by the information investors have[5]. This information is built on their knowledge about the technology behind cryptocurrencies, changing regulations that effect the types of cryptocurrencies (e.g. ICO, STO) and the exchange markets. New information about these factors affect the traders' behavior, which results in a highly short- and long-term volatile market with high opportunities and risks[6].

One of the main sources for information and therefore the driver for their beliefs are Twitter messages and their sentiments. Most publications correlate the polarization in social networks with price movements[7,8,9].

It is hypothesized that messages from the cryptocurrencies-foundations Twitter channels have an impact on the network sentiment and hence are drivers for market polarization. Therefore, this research analyzed single tweets from the cryptocurrencies-foundations Twitter channels and their effect on price movements. Derived from the presented aspects, this work formulates the hypothesis:

---

[3] *Nakamoto* (2008)
[4] *Buterin* (2013)
[5] *Fama* (1970)
[6] *Conrad* et al. (2018)
[7] *Georgoula* et al. (2015)
[8] *Lamon* et al. (2018)
[9] *Li* et al.

*"Twitter messages from cryptocurrencies foundations have an impact on price movements within 15 minutes."*

According to the hypothesis, it is possible to predict future exchange rates, which holds the opportunity for beneficial intraday trading. The major applications of this work are the abilities to filter beneficial Twitter messages and execute the trade as quickly as possible. In order to outperform other market participants a trading algorithm is being developed. The algorithm collects data from Twitter, analyzes the tweet, decides whether to buy or sell a cryptocurrency and the time of execution.

Nevertheless, to create a meaningful analysis, it is necessary to have a deep understanding of cryptocurrencies in order to identify potential drivers of their value and filter profitable information. Especially important aspects are the technological, typological and regulatory factors affecting the cryptocurrencies over time. Moreover, the practical application requires knowledge about the exchange markets and their conditions. After laying the fundamentals, the work moves forward to statistical research, where tweets are clustered under certain conditions. The return and excess return of these clusters are then analyzed for a period of 15 minutes after publication to assess the magnitude of value movements. Afterwards, the results are implemented to a basic trade algorithm and lastly, all of the findings are discussed.

## 2 Cryptocurrencies

A cryptocurrency, also known as a digital or virtual currency, has no central definition. However, most definitions are similar and contain the same aspects. The Oxford Dictionary for example defines a cryptocurrency as "a digital currency in which encryption techniques are used to regulate the generation of units of currency and verify the transfer of funds, operating independently of a central bank"[10]. According to the Merriam Webster Dictionary, a cryptocurrency is "any form of currency that only exists digitally, that usually has no central issuing or regulating authority but instead uses a decentralized system to record transactions and manage the issuance of new units, and that relies on cryptography to prevent counterfeiting and fraudulent transactions"[11]. Both definitions include the same core elements; Firstly, they are currencies, which are neither fiat nor commodity currencies, as they are without intrinsic value and monetary authority[12]. Secondly, they exist in a digital form.

---

[10] *Oxford Dictionaries*
[11] *Merriam-Webster*
[12] *Baur* et al. (2018)

Thirdly, the regulation of transactions is decentralized and lastly, they use encryption/cryptography techniques to ensure security for peer-to-peer trading and counteract the double-spending problem, where account-holders are able to bypass security and spend their electronic cash twice[13].

The rationale behind decentralization of the transaction regulation is proposed by Nakamoto as a termination of the trust-based model in financial institutions and the removal of third party intermediaries[14]. Moreover, the architectural decentralization decreases the political influence and control on the currency. The blockchain and the distributed ledger technology enable the possibility to create a decentralized network that algorithmically verifies transactions.

There are several debates whether cryptocurrencies are mediums of exchange as the name suggests, or speculative assets. In the current form, cryptocurrencies appeal to users less as a transaction system rather than an additional investment asset[15]. Initial coin offering (ICO), for example, is a utility token which uses the crowdfunding method to raise capital for a new cryptocurrency venture[16]. From an investors' point of view coin offerings and cryptocurrencies in general are highly speculative investments because their proposition is similar to fiat currencies that have no physical value underlying. However, there are additional types of cryptocurrencies, namely security and equity tokens, which marginally differentiate from utility tokens. Current cryptocurrencies are minimally regulated. However, it's worth noting that the landscape of cryptocurrencies is in constant flux, with many new entrants gravitating toward a level of regulation akin to that seen in traditional stock markets. In this context, altcoins may not significantly differ in terms of the technologies proposed by their respective foundations. Nonetheless, what sets them apart are the diverse advantages and disadvantages they bring to the table, each offering unique potentials for investors. As such, investors should exercise due diligence in assessing the specific characteristics, risks, and rewards associated with different altcoins before making investment decisions.

as higher regulations seek for additional functions.

---

[13] *Nakamoto* (2008)
[14] *Nakamoto* (2008)
[15] *Baur* et al. (2018)
[16] *Fisch* (2019)

## 2.1 Types of cryptocurrencies and pre-sales

Since the launch of Bitcoin, thousands of other cryptocurrencies were introduced to the ecosystem. As mentioned earlier, cryptocurrencies are based on different technologies, which results in different groups of currencies. The two major technological groups are coins and tokens. A coin operates independently and therefore has its own distributed ledger technology. A token, on the other hand, works on an external platform such as Ethereum, Omni and others[17]. Tokens are mostly based on smart contracts with the ERC20 technical standard. Nevertheless, the terminologies coin and tokens are usually used equally regardless the structural differences between them.

From a functional point of view, there are different kinds of tokens, the most important ones being utility, security and equity tokens. Utility tokens primarily serve the purpose to give access and functionality to a system or network[18]. However, the regulation standards are very low, which is permissive for illegal activities (e.g. fraud[19], pump and dump[20]). The migration of assets to the blockchain, which are evaluated by security regulations (e.g. Howey test), are called security tokens[21]. They represent a combination of cryptocurrencies and traditional stocks. Their value is derived from external and tradeable financial assets. However, security tokens require regulated exchange markets, which currently rarely exist. In the coming years, the promotion of the different exchange markets for security tokens are likely to attract high shares of Wall Street's money[22]. Moreover, security token offerings are predicted to end the ICOs era and increase the professionalism in the cryptocurrency market. The newest form of cryptocurrency token is the Equity token which is a subcategory of a security token. An equity token is similar to traditional assets as it implies shares of ownership[23].

The different types of coins are mainly used to raise capital by selling newly created tokens to a crowd of investors[24]. Analogue to an Initial Public Offering (IPO), cryptocurrency markets use Initial Coin Offerings (ICOs), Security Token Offerings (STOs), or Equity Token Offerings (ETOs) to promote their assets. The different kinds of offerings constitute the same process with different types of tokens underlying. They share high similarities to crowdfunding but are based on the distributed ledger technology[25]. Over 1,000 ICOs that were published on Coinschedule raised total funds of over $20B in 2018. The most successful ICO that was published so far is EOS. This ICO raised a total amount of over $4B in less than a year. However, the hype for ICOs has drastically decreased throughout 2018, as ~$17B of the total of $20B raised, were collected in the first half of the year[26]. From an investors point of view ICOs are highly speculative and hold high potential and risks. The high risk is especially reflected by the fact that most ICOs are initialized by start-ups with innovative ideas as well as fake companies, and only 22% of the ICOs launches in the third quarter of 2018 were successful[27]. The improvement of the regulation standards for STOs and ETOs is likely to drastically decrease these risks and increase the expertise in the cryptocurrency market. The driving factors for a successful pre-sale phase of ICOs greatly differ from IPOs. When ICOs are considered, the white paper, technological competencies and social media activities improve the amounts of funding raised[28]. The white paper primarily describes the technology, the concept, the intention and gives other information about the project. The technological competencies are the source code and relevant updates which are transparent and usually published on GitHub. The social media activities, most importantly Twitter, are the third factor that increase the success of an ICO[29]. These driving factors can be summarized to a global factor called public relations. Recent study show that price movements in the cryptocurrency market highly correlates with social media platforms and especially Twitter[30]. A Twitter message can contain news about all the technological, conceptual and other factors mentioned earlier. Thus, these messages can cover all the different dimensions and drive the value of cryptocurrencies. After the pre-sale phase, established altcoins can be added to the portfolio of different exchange markets to be further traded.

---

[17] *Momtaz* (2018)
[18] *Rennock* et al. (2018)
[19] *Elendner* et al. (2016)
[20] *Hamrich* et al. (2018)
[21] *Jones* (2018)
[22] *Koverko/Housser* (2017)
[23] *Fisch* (2019)
[24] *Fisch* (2019)

## 2.2 Cryptocurrency Exchange Markets

Cryptocurrencies are mainly traded on exchange platforms that are similar to traditional stock exchange markets. Traders can place buy and sell limit orders and the broker platform matches buyers and sellers when the conditions match. Unlike stock exchanges, cryptocurrency exchanges are permanently available and since they perform online, there are no demographic limits. The low demographic barriers in addition to the simplicity to create a new wallet can increase the attractiveness for new traders and therefore increase the market liquidity. Moreover, trades are accomplished directly when a match is found.

---

[25] *Fisch* (2019)
[26] *Coinschedule*
[27] *Scde Ventures*
[28] *Fisch* (2019)
[29] *Fisch* (2019)
[30] *Li* et al.

However, as mentioned in the utility and security token context, the regulation standards are yet to be fully developed. The same applies for the exchange markets in terms of security, internal controls, market surveillance protocols, and disclosures[31]. These immature standards can result in high losses for investors due to hackers activity[32,33]. Therefore, it is of high importance to investigate the different exchange platforms in terms of security and the possible impact of regulation standards on the markets.

Over 220 exchange markets exist which do not differ only in their security standards, such that more aspects are to be considered[34]. Each market differs by the volume traded, amount of tradeable coins, transaction fees, user-interfaces, and the foundations country of origin. A high trade volume and thus a greater liquidity is an important factor for intraday trading. The 30-day trade volume ranges from around 25 million to 21 billion dollars. The top five exchange markets are Binance ($21B), OKEx ($17.6B), Huobi ($14.1B), DigiFinex ($12.3B) and ZB.COM ($11.3B)[35]. Binance, OKEx and Huobi offer around 400 different coin exchanges, while DigiFinex and ZB.COM offer around 100 each. Thus, the trading opportunities are significantly lower in DigiFinex and ZB.COM. The trading fees on Binance are generally 0.1%, but they can be reduced by 25% if the trader holds Binance Coins (BNB) resulting in a potential fee of 0.075%[36]. The fees for DigiFinex[37], Huobi[38] and ZB.COM[39] are 0.2% per trade. OKEx offers a volume based discount model, where they differentiate between buyer and seller. The fee for buyers ranges from 0.02% to 0.1%, if the 30-day trade volume exceeds 50,000 BTC. For a seller, the fee ranges from 0.05% to 0.15% under the same conditions[40]. Each trading platform offers an Application Programming Interface (API), which is essential to operate the trading algorithm.

---

[31] *Underwood* (2018)
[32] *Zhao*
[33] *Partz*
[34] *CoinMarketCap*
[35] *CoinMarketCap*
[36] *Binance*
[37] *DigiFinex*
[38] *Huobi Global* (2019)
[39] *ZB.com*
[40] *OKEx*

The five introduced exchange markets are currently the biggest in terms of trade volume. They offer the best overall conditions but slightly differ from each other. Binance and OKEx have the most similar conditions. Nevertheless, the transaction fees of Binance outperforms OKEx when the total amount of Bitcoin traded during the month is lower than 10,000 BTC. For the performed research, a large trade volume of data dissected to the shortest time points is crucial to get accurate results. The API of Binance provides candlestick data in 1-minute intervals for every exchange pair. Thus, the database of Binance was chosen to be the source of trading data for this research.

*Table 1: Exchange market data*

|  | **Binance** | **OKEx** | **Huobi** | **DigiFinex** | **ZB.COM** |
|---|---|---|---|---|---|
| **Trade Volume** | $21B | $17.6B | $14.1B | $12.3B | $11.3B |
| **No. Markets** | 406 | 415 | 382 | 102 | 92 |
| **Transaction Fees** | 0.1% - 0.075% | Buyer: 0.1% - 0.02% Seller: 0.15% - 0.05% | 0.2% | 0.2% | 0.2% |
| **API Interface** | Yes | Yes | Yes | Yes | Yes |

## 2.3 Market volatility

Being a system in development, the cryptocurrencies hold high investment potential and risks. The previously mentioned technological and regulatory uncertainties increase the markets long-term volatility for cryptocurrencies over the stock market. The uncertainty and missing underlying assets reveal that the investments are mainly based on belief. In accordance, the long-term 30-day volatility chart for Bitcoin presents a decreasing trend, which according to the 7$^{th}$ of January 2019 had a standard deviation of daily returns of nearly 5%[41]. Nevertheless, the standardization progress of the regulatory perspective and the raising acceptance of the cryptocurrencies in the ssociety has a positive impact on the long-term volatility. The reduced volatility implies for the stabilization of Bitcoin and is of high importance since Bitcoin is the major currency that holds over 50% of the total market[42]. Most of the altcoin pairs are traded with Bitcoin and not fiat currencies. Therefore, Bitcoin has a high impact on the USD value and the value of altcoins increase or decrease according to the value of Bitcoin[43].

However, in addition to the long-term stabilization, the short-term volatility is of high importance for intraday trading in order to have a higher trading impact. Most trading strategies in the cryptocurrency market are based on speculative short-term volatility and algorithmic trading based on technical analysis rather than fundamental trading strategies like portfolios. Therefore the classical risk and return measures are not suitable to explain the cryptocurrencies' price movements. Instead, behavioral anomalies are shown to have a high relevance in the cryptocurrency market[44]. The most important among these anomalies is the overreaction bias, where investors overreact to price declines and raises[45]. The increased spread caused by behavioral anomalies holds a higher potential to gain profits. In combination to the overreaction phenomenon, the information spread has a significant influence on the price development. As there is no real underlying asset and the market is driven by uncertainty and belief, any source of information can have an impact on the cryptocurrencies' value. In their paper, Garcia and Schweitzer, demonstrated the impact of information spread via Twitter messages on price movements[46]. However, their analysis was based on general market polarization and sentiments and therefore on the mass opinion within the social media. Nonetheless, the public opinion is most likely to evolve and be derived from a specific environment. A possible explanation that forms such environment and thus the public's opinion about cryptocurrencies is the information given by foundations and experts which spread fast through the social media platforms. Therefore, identifying the messages that have a significant impact on the traders' opinion is of high importance to forecast price movements and take advantage over the high short-term volatility..

---

[41] *Buy Bitcoin Worldwide* (2018)
[42] *CoinMarketCap*
[43] *Ciaian* et al. (2018)

## 3    Research Model

The research model was developed based on two major conditions. First, the statistical relevance to the hypothesis *"Twitter messages from cryptocurrencies foundations have an impact on price movements within 15 minutes."* and second, the possibility to apply the results to the trading algorithm. In order to identify the impact of Twitter messages on trading behavior, two datasets were used. Historical Twitter messages and trading data from binance. The Twitter API provides the GET function "user_timeline", which returns the latest up to 3,200 Tweets from a specified user[47]. This function was used to obtain the latest 3,200 Tweets from the altcoins official channels that are listed on Binance. Each dataset has 88 columns that contain additional information about the Tweet (e.g. timestamp, hashtags, retweet). In total, the raw Twitter timeline data comprises 136 timelines with 147,111 messages. The second dataset, the historical trading data, was obtained via the API of Binance. They offer the GET function "klines", which returns open, high, low, close and volume indicators (OHLCV) in 1-minute-intervals for all of their listed pairs. The retrieved datasets ranges from 2 January 2018 (Unix 1514851200) and 10 November 2018 (Unix 1541889720) and contain 450,000 observations per cryptocurrency. Only pairs that are converted to BTC are considered in the analysis in order to exclude the Bitcoins high influence on the altcoins USD value. The logarithmic return based on the 1-minute-open-intervals is then used as an indicator of the price development.

---

[44] *Yang* (2018)
[45] *Chevapatrakul/Mascia* (2018)
[46] *Garcia/Schweitzer* (2015)
[47] *Twitter*

Both datasets are combined in a standardized manner, where a Twitter message from a cryptocurrencies timeline is linked to the corresponding return rates. The creation time of the Twitter message is the starting point of standardization. $T_1$ is the timestamp of the return rate for the first rate of return that follows the creation time of the Twitter message, where the difference between the creation time and the binance timestamp is 0 seconds $\leq t_1 < 60$ seconds. Fifteen return rates for $t_1 - t_{15}$ illustrate the 15-minute price development after each Twitter message. However, only full datasets are considered i.e. Twitter messages that are not older than the available historical trading data and have no missing trading data. Some trading data is missing due to server maintenances, where no data was collected[48]. After the standardization process, 55,911 full observations remained.

After the data preparation, the remaining dataset was conditionally analyzed. This means that only Twitter messages are considered that fulfill specified conditions. However, when only conditional variables are considered that occur more than $n \geq 50$ times, the 88 different Twitter variables hold over $10^{41}$ possible conditions. The high amount of conditions required too much computational power to be fully analyzed and therefore a reduction of dimensions is required. Out of the 88 variables, about 70% were deleted, as they had no influence on this research (e.g. URL variables, coordinate variables) and variables that can be derived from other variables (e.g. Boolean variable is_retweet can be derived from retweet_text). After the first deletion of obvious variables, two correlation matrices were created to reduce additional dimensions. The first correlation matrix measured the correlation between Boolean variables, where TRUE means that the variable is unequal to NA and FALSE means that the variable is NA. With this correlation matrix, variables were identified that occur together.

---

[48] *Reddit*

Variables with a correlation of one that can be derived from another variable were therefore deleted. The second correlation matrix contained the raw data of the variables. This identified more potential of variables that could be deleted. Created_at has a high correlation with quoted_created_at (0.949) and retweet_created_at (0.971). Moreover, followers_count and listed_count correlate with 0.932. Thus, the three variables quoted_created_at, retweet_created_at and listed_count were deleted. After the reduction of dimensions, 19 variables were left (see appendix A). In addition to the variables provided by Twitter, the sentiment for each Twitter message was added as an additional variable. It is the average sentiment from the sentiment_by function with default settings of the sentimentr package[49] for R.

Afterwards, the leftover variables were prepared for the conditional analysis. Numeric variables were clustered to five equally ranged groups between minimum and maximum. Text variables were separated to single words and stop words were deleted with the tidytext Package of R[50]. During the conditional analysis, the 52,911 Twitter messages were iteratively filtered by all possible variable combinations. For each combination, the amount of leftover observations n, minimum and maximum, 1st and 3rd quartile, mean, median and standard deviation for $t_1 - t_{15}$ were calculated. For the data analysis, only sample sizes of n ≥ 50 were taken under consideration. The amount of observation points not only accounts for the representability but also the amount of trades for the trading algorithm. Fifty observations means an average of around one trade per week.

Afterwards, the same research model was applied to the excess return rates instead of the return rates. This additional method validates whether the Twitter message had a positive impact over the current momentum of the coin. For the excess return rate, three possible market models were taken under consideration. The first two options are based on the capital asset pricing model (CAPM)[51] without consideration of the risk free rate. For the first possibility, the market return $r_m$ is the Bitcoin to USD ratio due to the high market influence the Bitcoin has. However, the model shows a low coefficient of determination as the altcoins exchange is Bitcoin, while the Bitcoins exchange is USD. Thus, the additional currency-exchange rate has an influence on the price movement. Moreover, the high market volatility of the minutely trading data has a negative effect on the explanatory power. For the second option, the market return $r_m$ is the average of all cryptocurrencies that can be exchanged to Bitcoin on Binance. For some cryptocurrencies, this approach reveals potential with partial correlations of over 0.5 for minutely returns. However, the average correlation is 0.16 and therefore the explanatory power is too low to be used as market model. Due to the high volatility, the third option is a simple moving average with the previous n = 30 (30 minutes) returns. The amount of return rates considered in the simple moving average is a trade-off between the current momentum and volatility. A higher amount of return rates reduces the volatility and weakens the momentum, while a lower amount increases the volatility and increases the momentum. However, this model sets the focus on the cryptocurrency of investigation, catches the momentum of the altcoin, and reduces the impact of outlier values. The 19 different variables can be clustered to two major groups, text conditions and exposure rate. The sentiment, word used in the message, media type and language are factors that represent the text condition. The screen name, followers count, friends count, statuses count, favorites count, hashtags and mentions screen names are indicators for the exposure rate. The sentiment is a numeric value for text analysis that ranges from -1.196 to 1.534 in this research. A negative number reflects a negative and a positive number reflects a positive polarity, neutral messages are around zero. The polarity of the message might be an indicator whether the price increases or decreases. A word used in the message in addition to the sentiment of the text specifies the Tweet. For example, a message about the technology or a new partnership with a positive or neutral sentiment might have a higher influence on the trading behavior. A tweet that contains a media type (only photos) is unlikely to have a positive influence when grouped as single condition. However, a photo in combination with other factors can have an impact on the intention of the message (e.g., pictures are mostly entertaining or informative). The used language mostly describes the text condition but further contains exposure elements. English messages are most likely to be understood by everyone in the network, while different languages reduce the exposure rate and thus the amount of people that possibly react to the message. However, the language can reach different audiences and in accordance to the home bias[52] investors might overreact to messages in their mother tongue. The sentiment analysis is based on an English dictionary and thus the analysis can provide wrong results for other languages. Moreover, the language is an indicator for the exposure. Besides for the used language, the screen name is a summary of the exposure factors followers, friends, statuses and favorites count. In addition to the numeric factors, the quality of messages and therefore the public influence can differ between Twitter accounts. The followers count is a value that directly indicates the interest of people in the cryptocurrency. Hence, the amount of traders influenced by a message might increase accordingly, if the intentions of each person following the account are equal. The friends count is a value that returns the amount of users the specified account is following. It might show the activity and connections within the network. However, it is most likely that this factor has no influence. The statuses count is a measure for tweets issued by the user. It is a direct indicator for the activity within the network. It might illustrate whether the accounts messages are higher quality or quantity. Favorites count reveals the number of tweets this user has liked. Analogue to the friends count it expresses the activity of the user but is most likely to have no influence on the trading behavior. The hashtag function is another method to control the exposure of a tweet within the network. People that follow a specific hashtag can see the message without specifically following the account. Hence, hashtags can increase the exposure rate in the network. Correspondingly, mentioned users in a tweet spread the message through timelines from users that follow the mention and thus increases the exposure rate. The retweet and quote function have two possible outcomes. On one hand a retweet means that the message has a higher exposure rate as the message is spread and therefore it might have a high importance. On the other hand, the existence of the message means that there is a time disadvantage. Thus, the intention of the trading algorithm to have a first mover advantage is in danger. However, the friends and favorites count for retweets and quotes give additional information whether a high exposure rate has a positive or negative influence. This research methodology holds three major advantages. First, the conditional analysis reveals the interplay between different factors. Second, the conditions can be implemented to the trading algorithm without additional calculations and therefore use minimalistic computational power to gain speed advantages. Third, this form of standardized processing can be used to optimize buy and sell time points.

---

[49] *Rinker* (2018)
[50] *Silge* (2018)
[51] *Sharpe* (1964)

---

[52] *Tesar/Werner* (1995)

## 4 Results

Cryptocurrencies were going through a hard year in 2018. After the peak at the end of 2017, where one BTC nearly hit the value of 20,000 USD, the market collapsed in 2018. During the period of investigation from 2 January 2018 to 10 November 2018, the BTC price deflated from 14,773 USD to 6,403 USD and lost around 57% of its value (see figure 2 left side). The price development reveals that the volatility decreased over time and the market stabilized at the second half of the year. However, the cryptocurrency to USD development has minor importance for the trading algorithm itself as it is based on the altcoin to BTC ratio. Nonetheless, an increase of BTC is not beneficial if the USD decrease is higher than the BTC increase.

A closer look at the value of five altcoins to BTC during the period of investigation reveals similarities to the BTC-USD development (see figure 2 right side). Prices highly decreased on the long-term view. The average loss for all cryptocurrencies investigated in this research was -54% with a standard deviation of 41%. The best performance was Binance Coin (BNB) with a price gain of 139%, while Insolar (INS) lost over 99% of its value. The graphs show that although most cryptocurrencies show a negative trend in average, it is possible to take advantage over the upstreams to increase BTC.

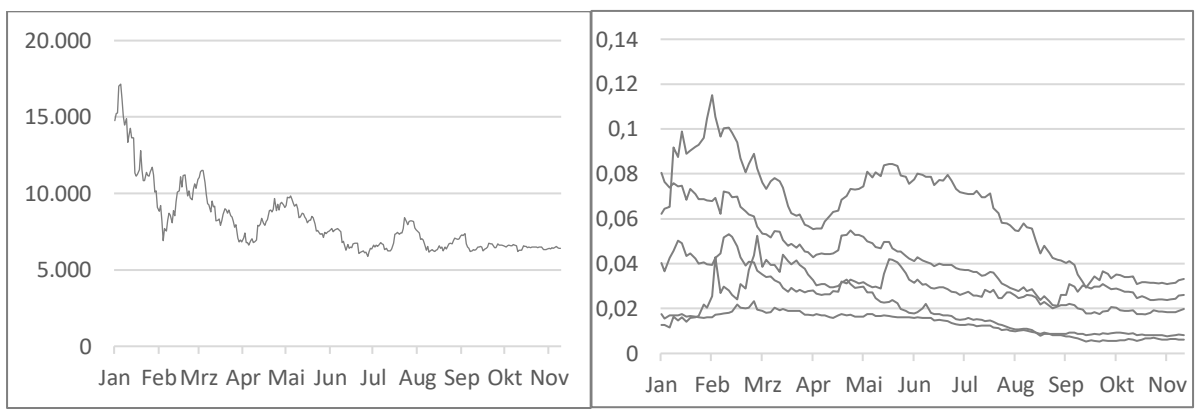

*Figure 1: Price development BTC-USD (left) and the altcoins DASH, ETH, ZEC, LTC and DGD to BTC (right) from 2 January 2018 - 10 November 2018*
*Source: Both graphs are based on Binance data*

The average sentiment of the Twitter messages is 0.139 with a minimum of -1.196 and a maximum of 1.534 and a standard deviation of 0.162. The sentiments are normally distributed around the average with a single peak at zero. Hence, most messages that are published on Twitter have a neutral sentiment. However, the average slightly over zero shows that although prices drastically deflated the general sentiment is positive. The positive sentiment results due to the fact, that all the Twitter messages are from foundations of cryptocurrencies and thus most of their Tweets have a positive consensus. Over 82% of the words that were tweeted were taken under consideration (1,250,578 out of 1,516,484 words). The other part of the words are ignored because they are stop words and therefore most likely irrelevant. Moreover, words that are used less than fifty times were excluded. Most of those words were tweeted in a different language than English and therefore the probability of occurrence is low. Nonetheless, the majority of the messages with over 95% are recognized in English and 2% are undefined (mostly posted links). The different languages are spread over all of the different Twitter accounts. Most foundations write parts of their messages in their mother tongue. A closer look at the message quantity reveals that there are significant differences in the message activity of the different accounts varying from a minimum of one to a maximum of 2599 Tweets during the period of investigation. The top three accounts are eth_classic, KomodoPlatform and NEMofficial. They have sent over 2,000 messages, which accounts for an average of eight Tweets per day. However, the amount of Tweets seems to have no influence on the followers count as the correlation is only around 0.1. Hence, traders are attracted by different factors than the presence in the social media. Most likely, the quality of the posts has an impact and thus the followers count can be interpreted as a quality indicator. The followers count for most cryptocurrencies is relatively low. 91 out of 118 Twitter-IDs have a follower's count of less than 100,000. Binance (878,560) and Ripple (906,430) got most followers by far. Litecoin (433,085) and Ethereum (433,530) are ranked three and four but have half of their followers, although their market capitalization is quite similar. The distribution reveals a high range difference between the accounts and therefore their exposure rate. The network activity indicators friends and favorites count reveal high differences through the accounts. However, they only have a low influence on the followers count. Although the linear relationship between most of the numeric variables is low, the amount of Twitter messages and the favorites count shows a moderate correlation (0.46). Both factors can be summarized as messaging activities, while the friends and followers activities can be summarized as networking activities. The hashtag function is used in 28.5% of all Twitter messages in the dataset. The top hashtag is #blockchain, which is used in nearly a quarter of times out of all messages including a hashtag. It is followed by #crypto and #cryptocurrency, which together account for nearly 17%. They are cryptocurrency specific but global hashtags that have a high exposure rate. They can be meaningfully used by all the Twitter accounts. Other hashtags are coin specific (e.g., #BTC, #Ethereum) or general hashtags that potentially reach mainstream users (e.g., #news, #trading). Therefore, all of the hashtags possibly reach different customer segments. In addition to hashtags, the mentions function is used in a similar manner (28.5% out of all messages). The person with most mentions is Justin Sun (@justinsuntron) with slightly over 5%. However, in contrast to the hashtags the distribution of the mentions is minor. The three major mention groups are founders and experts (e.g. @justinsuntron, @MattSpoke), cryptocurrencies (e.g. @ethereum, @KomodoPlatform) and exchange platforms (e.g. @binance, @bitfinex). Each of these mention-groups hold information potential for different aspects. Founders and experts as well as cryptocurrencies have inside knowledge and thus possibly give important information about technology changes, new partnerships et cetera. Messages connected to exchange platforms on the other hand might reveal relevant trader-specific information. Besides hashtags and mentions, an additional way to spread information are the quote and retweet functions Twitter offers. Both functions are used in nearly one third of all messages investigated. However, both functions do not work together. Out of all messages that used retweets and quotes, Retweets are used over 80% of the time and therefore have a higher importance.

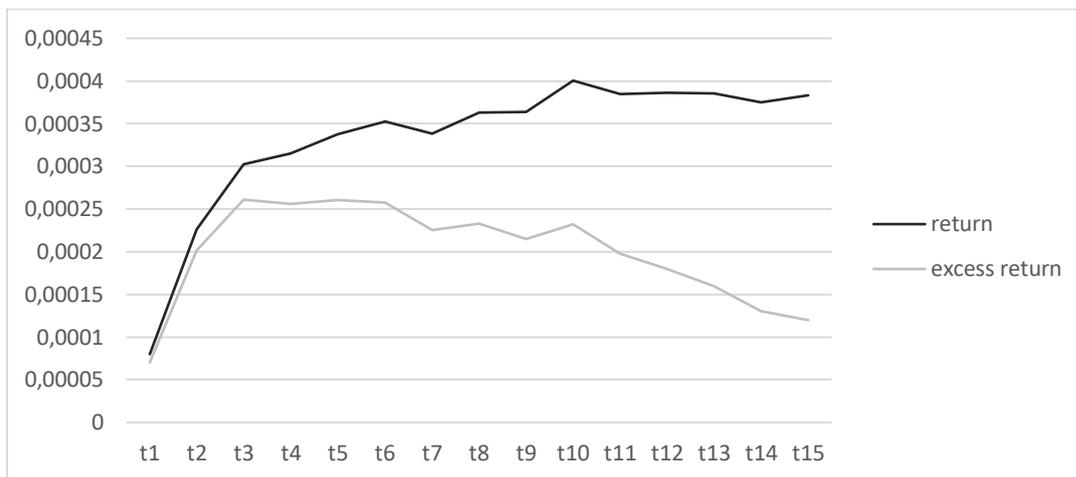

*Figure 2: Cumulative return vs. excess return price development 15 minutes after Tweet was published*

Overall, Twitter messages generally seem to have a high impact on the price movement directly after the Tweet was send. Figure 3 shows the cumulative average return and excess return for all investigated Twitter messages 15 minutes after the Tweet was send. The trends for the returns as well as the excess returns show a very high slope from $t_1$ to $t_3$ right after the Tweet was send. This indicates a high short-term impact right after the messages were published and thus very short reaction timings. In addition to the immediate response, the returns show a slightly positive trend until 10 minutes after. However, taken under consideration the price development for all cryptocurrencies of -54% presented earlier, signals from the foundations have a positive influence. Nonetheless, that the excess return is lower than the return rate shows that the considered cryptocurrencies had a slight uptrend during the period the tweets were send. The downtrend of the excess return is due to the disadvantage of the methodology for the excess return rate calculation. As the time period of the moving average changes towards $t_{15}$ the impact of the positive price movement increases the value and thus decreases the excess return rate. However, both graphs show a high impact of the Twitter messages right after they were published. The return rate maximizes at $t_{10}$ with a cumulative return of 0.04%, which is too low to be valuable considering the transaction fees of 0.075%.

However, in addition to the unambiguous price movement shown in figure 3, a paired two-tailed t-test between all the returns of the fifteen time periods after a tweet was send give clarification about the significance of the result. The p-values shown in table 2 (next page) reveal the significance levels between the different phases. Especially the three periods right after the Tweet was published reveal high differences in comparison to the other phases. They are significantly different to all phases except for $t_{10}$. However, $t_2$ shows the highest differences to all the other phases. While the p-value is smaller than 0.05 for $t_1$ and $t_3$, it is smaller than 0.0001 for the rest. Moreover, $t_2$ has the highest average return with 0.000146, which is nearly double as high as it is for $t_1$ and $t_3$. The other periods $t_4$ - $t_{15}$ average slightly around zero and show mostly no significant differences between each other. In conclusion, Twitter messages show an immediate reaction, which significantly peaks at $t_2$. Moreover, the market gets back to the normal state after $t_3$, as $t_4$ to $t_{15}$ predominantly show no significant differences. However, the standard deviation reaches the maximum at $t_2$ with 0.004384, which shows that there are high differences between the impacts of each message. Nonetheless, all of the results presented until now reveal that certain messages have a significant impact on traders' behavior but requires a deeper analysis about the influence of different factors in order to identify message clusters with high potential.

In contrast to the influence of the sentiment of all messages within the Twitter network,[53] the sentiment of specific messages seems to have no influence on the impact of a message. For all the sentiment groups the price movements are similar to the return presented in figure 3 and hence are slightly above zero for $t_1$ to $t_{15}$. However, the group with a sentiment of -0.104 to 0.442 and thus the mostly neutral messages have a marginally higher return with a lower volatility compared to the other groups. This might be due to the fact, that neutral messages are mostly informative. Tweets with a sentiment greater than 0.6 are mostly replies (60%) to other users, specifically retweets (31%) or quotes (29%) and hence interactions with other users, while tweets with a sentiment lesser than -0.1 are mostly reactions to problems like frauds or other issues.

The price movement for clusters by screen name reveals high differences and potential. None of the Twitter messages have a negative effect on the price movement, as the foundations do not post messages that damage their own cryptocurrency. However, there are differences between tweets that have no or positive influence. Especially users with a relatively low Twitter activity seem to have higher return rates. The maximal cumulative return during the

---

[53] *Li* et al.

15-minute period is 0.0017 for group 1 with less than 200 messages, 0.0011 for group 2 that tweets more than 200 and less than 500 times and 0.00045 for group 3 with more than 500 tweets. The p-value for group 1 and 2 is 0.1 and has no significant difference. However, group 1 and 3 as well as group 2 and 3 are significantly different with $p < 0.01$. The message activity can therefore be interpreted as a quality factor. Users that tweet less are more likely to publish informative and relevant messages, while users with a higher activity tend to answer user questions et cetera that have no relevance to the price of the cryptocurrency.

*Table 2: T-Test matrix for the returns of the fifteen periods after Tweet*

| ROI | $t_1$ | $t_2$ | $t_3$ | $t_4$ | $t_5$ | $t_6$ | $t_7$ | $t_8$ | $t_9$ | $t_{10}$ | $t_{11}$ | $t_{12}$ | $t_{13}$ | $t_{14}$ | $t_{15}$ |
|---|---|---|---|---|---|---|---|---|---|---|---|---|---|---|---|
| $t_1$ | - | | | | | | | | | | | | | | |
| $t_2$ | .023* | - | | | | | | | | | | | | | |
| $t_3$ | .904 | .018* | - | | | | | | | | | | | | |
| $t_4$ | .007** | .000*** | .022* | - | | | | | | | | | | | |
| $t_5$ | .026* | .000*** | .045* | .728 | - | | | | | | | | | | |
| $t_6$ | .013* | .000*** | .019* | .932 | .831 | - | | | | | | | | | |
| $t_7$ | .000*** | .000*** | .000*** | .282 | .170 | .264 | - | | | | | | | | |
| $t_8$ | .029* | .000*** | .038* | .622 | .904 | .810 | .173 | - | | | | | | | |
| $t_9$ | .002** | .000*** | .003** | .628 | .403 | .504 | .543 | .397 | - | | | | | | |
| $t_{10}$ | .097 | .000*** | .132 | .331 | .555 | .405 | .034* | .536 | .167 | - | | | | | |
| $t_{11}$ | .000*** | .000*** | .000*** | .260 | .129 | .183 | .925 | .098 | .481 | .046* | - | | | | |
| $t_{12}$ | .002** | .000*** | .003** | .654 | .424 | .566 | .498 | .402 | .930 | .165 | .484 | - | | | |
| $t_{13}$ | .001*** | .000*** | .004** | .592 | .399 | .472 | .573 | .335 | .971 | .117 | .515 | .913 | - | | |
| $t_{14}$ | .001*** | .000*** | .000*** | .359 | .204 | .269 | .905 | .154 | .638 | .058 | .831 | .585 | .708 | - | |
| $t_{15}$ | .004** | .000*** | .008** | .860 | .598 | .708 | .352 | .519 | .760 | .211 | .304 | .837 | .737 | .490 | - |
| **Mean** | .000080 | .000146 | .000077 | .000013 | .000023 | .000015 | -.000014 | .000025 | .000000 | .000037 | -.000016 | .000001 | -.000001 | -.000011 | .000008 |
| **SD** | .004165 | .004384 | .004264 | .004054 | .004222 | .004280 | .004104 | .003964 | .004025 | .004169 | .004062 | .004201 | .004179 | .004277 | .003977 |

*Significance levels: *<0.05, **<0.001, ***<0.0001, p-values are calculated via paired two-tailed t-test.*

The statuses count reflects the amount of tweets per user since the creation of the account and as mentioned before can be seen as a quality factor. A closer look at the 15-minute price movement reveals the impact of the Twitter activity (see figure 4). The graph shows four groups with a different amount of tweets ranging from zero to 5072 messages since creation. Especially the messages for the two smaller groups show an immediate impact from $t_1$ to $t_3$ while for the two bigger groups the tweets seem to have no effect on the price movement. However, messages in the group with the lowest activity reveal the highest influence on traders' behavior but does not hold enough potential to be beneficial. As clustering via screen name seemed partially beneficial, most likely additional factors have a positive impact as the screen name groups multiple factors. In addition to the quantity of the tweets, the followers, friends and favorites count can be seen as scope factors of the foundations Twitter channel. However, the three scope factors seem to have no direct influence on the price movement. Nonetheless, the cluster with the highest favorites count between 5,775 and 8,050 has a high impact on the price movement reaching a maximum cumulative return of 0.0016 at $t_9$. It remains questionable whether the favorites count has a positive impact on the price movement or is coincidental. The cluster contains only three different Twitter users vergecurrency, SubstratumNet and GroestlcoinTeam, which might be connected to other factors. However, if the favorites count is coincidental and the scope factors have no impact on price movements, this means that there are only a handful of professional traders that immediately react to Twitter messages and they are present in most relevant channels. Moreover, the other part of the followers do not react to messages or react with a lag higher than 15 minutes. In addition to the scope factors of the foundations Twitter channel, the hashtag and the mentions functions hold potential to reach people who not purposely follow the specific cryptocurrency. However, messages that use hashtags compared to messages without hashtags show the same price movement. Moreover, no specific hashtag has a positive effect. Hence, traders that immediately react to Twitter messages follow quality sources already and other users reached by the hashtag do not either react or react with a lag greater than 15 minutes. Tweets that used the mentions function show like the hashtag function no significant difference to price movements. In addition, all the specific mention screen names except for Binance had no influence. It remains questionable, whether the scope or the content of the messages addressed to Binance have the effect. On one hand, the messages that include Binance refer to the cryptocurrency and therefore is an informative message. On the other hand, traders who use Binance are likely to track news about that exchange on Twitter. As the return data bases on Binance both scenarios can have an impact on the rates.

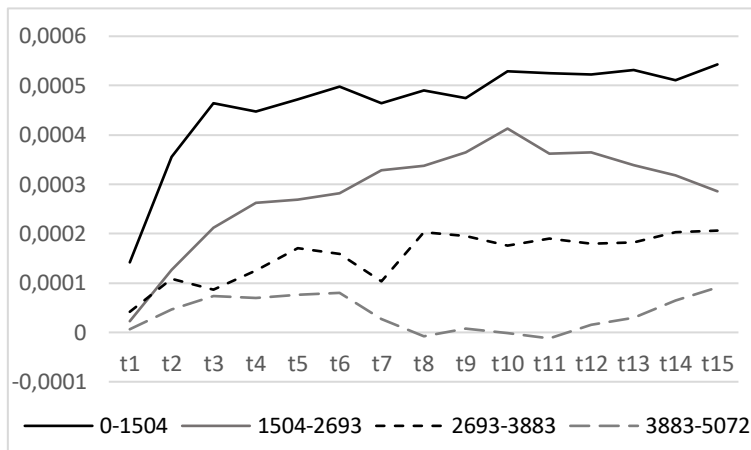

*Figure 3: 15-minute cumulative return after tweet for statuses count clusters*

Besides the quality and scope factors, some words used in the tweet seems to have an impact on the price movement. Especially words that imply technical or structural company changes like announce, partnered, launches, finally and cooperation reveal a maximal cumulative return of 0.005 within the 15 minutes range. Moreover, words that connect to exchange markets (e.g. binance, bitfinex), as pointed out with the mentions function, have a maximal cumulative return of 0.0043. Most of these messages are connected to an announcement that the cryptocurrencies are newly tradeable on these platforms. When the cryptocurrency joins a new exchange market, it suggests growth and increment of potential buyers. Moreover, it can be seen as a positive rating from the exchange platform. Nonetheless, many words in a tweet indicate no influence on price movements. Most importantly words that describe emotions or are without connection to changes of the company. In addition to emotional and unconnected words, pictures were found to have no influence. Thus, pictures are mostly used as an entertaining rather than an informative factor.

The different languages have no influence on traders' behavior. Therefore, overreactions do not appear due to the home bias as the amount of people these messages reach are probably too small. However, French messages are published by different Twitter channels and show a positive price development. In contrast to the other clusters introduced so far, French messages peak at $t_9$ with a cumulative return of 0.0027. It needs to be taken under consideration that the language recognition of Twitter is not always correct as the French and English languages share high similarities. Thus, English tweets are partially recognized as French tweets. Moreover, the observation only includes 110 tweets and therefore might be coincidental as no significance is given.

A comparison of all retweets, quotes and regular messages reveal differences. Retweets seem to have only a little impact on price movements from $t_1$ to $t_5$, reaching a maximal cumulative excess return of 0.0002. Since a retweet is just a republishing of an existing message, it contains no new information. Moreover, the messages have an average lag of 3 hours and hence traders most likely reacted to the information already. Nonetheless, there are high differences between the retweet lag ranging from seconds to days. When assuming a maximal retweet lag of 3 minutes, the message has a higher impact that results in a maximal cumulative return of 0.0008 at $t_5$. In conclusion, a retweet enhances the original message when the lag is short while it has no effect on price movements when lag is high, as the majority most likely reacted to the information already. Nonetheless, the followers and friends count as well as the retweet username have no influence on traders behavior. Traders are likely to trust in the source that made the retweet and do not react to the original source. However, the amount of words is too low in order to be meaningfully analyzed. In contrast to retweets, the additional information added to a quote seems to have an impact on traders' behavior over retweets and is similar to regular messages. Similar to retweets, quotes have a high lag variety between the original and the quote tweet. However, the average lag is 10 hours and therefore much higher than it is for retweets. When assuming a maximal lag of 3 minutes, the maximal average cumulative return is 0.0023 at $t_{12}$ and therefore much higher than it is for retweets. Like retweets, the followers and friends count as well as the origination of the message have no influence on price movements for similar reasons.

Overall, some Twitter factors considered in this analysis reveal high potential while others seem to have no impact on price movements at all. Especially the statuses count that reflects the quality of the tweets and the word used in a tweet that is an indicator for the direction of information were found to have the highest short-term impact on traders' behavior. Scope factors on the other hand like the followers or friends count have only a little immediate influence. However, taken under consideration the transaction fees and risks for most of the conditions revealed in this research 15-minutes trading seems not beneficial. Nonetheless, the fact that the cryptocurrencies show an immediate response after a tweet was published implies that some traders speculate positive long-term price movements. The group with the maximal return of 1.67% has a standard deviation of 4%. A trade is beneficial, when the min-max spread of the cumulative return is greater or equal than 0.5% and the excess return is positive during the 15-minutes period. Additionally, the minimal cumulative return must be between $t_1$ to $t_3$, otherwise the price increase is considered coincidental and has no connection to the tweet. An additional requirement is that the maximal cumulative return appears with a lag of at least five periods after the minimal cumulative return. After the application of all the requirements, there were 133 possible conditions for a Twitter message left that are considered beneficial. All of the conditions exclude retweets and quotes, while the focus is mostly on certain words used and a few Twitter channels. The prepared data including the preconditions and the optimal strike points were listed in a datasheet that is used in the trading algorithm.

# 5 Algorithmic Trading

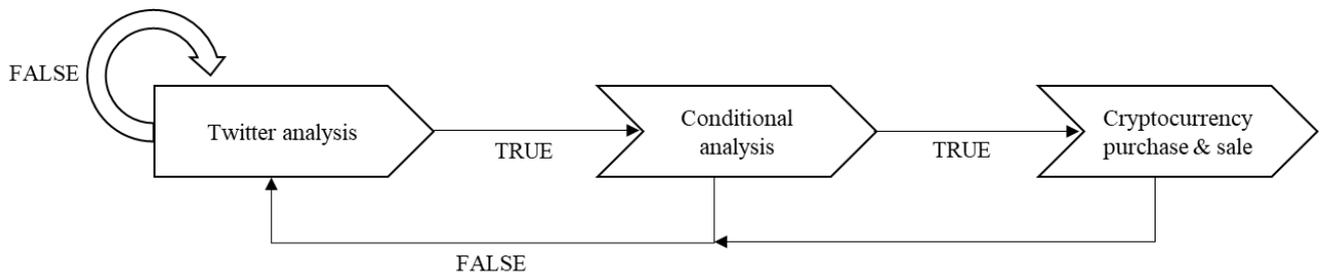

*Figure 4: Flow chart of the trade algorithm*

The trade algorithm as result of this work consists three different phases; the Twitter analysis, the conditional analysis and the cryptocurrency purchase and sale (see figure 5). The first phase is an infinite loop that continuously checks the defined Twitter channels for new messages. The conditional analysis phase compares the Twitter message with the results presented in chapter four and the last phase is the execution of the purchase and sale, assumed the tweet contains beneficial conditions.

The code of the trade algorithm (see appendix B) is based on R and uses four packages: rtweet[54], httr[55], jsonlite[56] and openssl[57]. Rtweet is used to ensure the communication with the Twitter API to retrieve user timelines, while the other three packages establish the connection with the Binance API. However, before the trade algorithm can work, several data needs to be loaded. A token for the Twitter OAuth authorization[58], the API and security keys for Binance[59] and two prepared datasets. The first database contains a list of the Twitter usernames and the associated Binance symbol identification in order to retrieve the relevant Twitter timelines and information about the cryptocurrencies for the purchases. Second, a prepared datasheet that as a result from the conditional analysis contains the relevant conditions as prerequisite for execution. Moreover, it contains the time for the minimal average cumulative return as purchase point and the maximal average cumulative return as sale point. In addition, the conditions are sorted by the min-max spread from highest to lowest to ensure that the most beneficial option will be chosen. The last variable to be defined is the spend rate that determines the amount of BTC in possession of the trader to be spent.

The first phase is an endless loop that retrieves the latest tweet from the user timelines defined in the imported sheet. The creation time of the message is then analyzed whether it is newer or older than

---

[54] *Kearney* (2018)
[55] *Wickham* (2018)
[56] *Ooms* (2018)
[57] *Ooms* (2019)
[58] *Twitter*
[59] *Binance*

60 seconds. If it is older than 60 seconds, the algorithm retrieves the next timeline of the list. If it was published less than 60 seconds ago, the algorithm moves to the second phase. A big disadvantage of the Twitter API is the download limit of 900 timelines per 15 minutes as a free user. Therefore, it is not possible to analyze the current timelines without limitations. One limitation is a stop timer that makes the code stop before every retrieve timeline loop. The sleep timer is defined as the time between the reset of the rate limit (900 seconds) divided by the rate limit (900) multiplied by the amount of user timelines to retrieve (variable). Hence, the timelines taken under consideration need to be optimized in order to maximize the outcome as tweets that are made during the waiting time might not be analyzed.

However, when the algorithm detects a new Twitter message, the code jumps into the second phase, the conditional analysis. All retweets and quotes will immediately end the loop as no condition was found where the trade would be beneficial. However, tweets that pass the first retweet and quote barrier are analyzed according to the conditions of the datasheet. If a match is found, the algorithm moves to the execution phase, otherwise it goes back to phase one.

The last phase is a rather static purchase and sale approach based on the results presented in chapter four that the algorithm uses. Before the purchase of the cryptocurrency, the code waits for the optimal point. When the optimal point is reached, the relevant information for the purchase is being gathered. The quantity to buy is the amount of BTC in the accounts wallet multiplied with the spend rate defined at the beginning. Afterwards, the amount to spend is divided by the lowest current ask price. The cryptocurrency to buy (symbol) is the match to the Twitter channel that published the tweet. However, Binance offers different purchase types (limit order, market order, stop loss, stop loss limit, take profit, take profit limit and limit maker)[60]. The algorithm presented in this work uses the market order that immediately executes at the current market price. This method ensures an immediate purchase execution without lag provided liquidity. After the purchase, the algorithm then the cryptocurrency until the time point of the maximal average cumulative return is reached within the pre-defined 15 minutes period.

Overall, the proposed trading algorithm is a basic approach that has further room for improvements. The 900-request limit from the Twitter API can be resolved by buying a premium account. Moreover, the sleep timer can be reduced by lowering the amount of timelines requested and therefore the amount of cryptocurrencies considered. A reduction of the sleep timer to less than 60 seconds will result in all Twitter messages being conditionally analyzed. Another possible approach is the increment of the threshold value. The second disadvantage is the static purchase and sell approach as an effect of the chosen methodology without respect to the current momentum. Combining the proposed trading algorithm with technical indicators can optimize the strike points and highly improve the outcome. Lastly, this thesis has only taken the 15-minute period under consideration without respect to long-term effects. An extent contemplation holds the opportunity for further insights and additional chances.

---

[60] *Binance*

# 6 Discussion

The research investigated the immediate impact of Twitter messages from cryptocurrency foundations on price movements in the Binance exchange market and implemented a basic trade algorithm based on the findings. A t-test matrix with the p-values has shown (see table 2), that the return rate for $t_1$ to $t_3$ is significantly higher than $t_4$ to $t_{15}$ for most comparisons. The price movement and p-values have unarguable proven the immediate impact of Twitter messages from trusted sources. Moreover, it was found that messages from foundations have no negative effect. Obviously, the investigated channels are not interested to harm their cryptocurrency and thus they do not publish negative tweets. However, the increment of the return rate is not high enough to be beneficial. Hence, only a low amount of traders track the messages published and react to them. Moreover, the partially 3 minutes lag until reaction and the high standard deviation in the grouped clusters show that the majority manually decides whether to invest or not. Additionally, the fact that certain wordings that suggest technological changes or new partnerships in a Twitter message increase the return rate to a beneficial level suggest that traders react to the given information. Provided a higher message quantity can be interpreted as reduced quality the statuses count is another indicator that suggests manual decision-making.

In contrast to the findings that the Twitter sentiment of tweets in the whole network[61][62], the sentiment of specific messages from foundations channels had no influence on price movements. The difference is in the perspective of the messages. The network or market sentiment is from the perspective of all potential buyers on Twitter while foundations messages are from the sellers' point of view. Buyers seek for information that sellers provide. The sentiment has no valuable role in a message, as it does not reflect the quality of the information content. In conclusion, quality factors like wordings and statuses counts are considered to have an impact on traders' behavior, while additional factors like hashtags and pictures were found to have no influence.

---

[61] *Phillips/Gorse* (2017)
[62] *Li* et al.

The scope factors like the followers or friends count as well as retweets and quotes have no significant influence on price movements. This strengthens the assumption only a low amount of traders show an immediate reaction within fifteen minutes after the tweet was send. Additionally, they are present in all of the channels, as there were no differences between different users. Nonetheless, retweets and quotes have no influence on price movements unless the lag is smaller than three minutes. It remains questionable whether the impact is from the original message or the new message. It is most likely that the positive effect is from the original message as the retweet is in range of the maximal slope. Moreover, retweets and quotes with a lag higher than three minutes have no influence on price movements and therefore traders' do not react twice to the same message. It remains questionable if quality messages have a long-term effect. A low amount of short-term reactions could mean that only professional traders' show an immediate reaction, while amateurs realize the news with a higher lag or from different media. Thus, it is important to investigate the long-term effect of Twitter messages as well as the effect of other channels and the publication timings between them.

The trading algorithm proposed in this work is a basic and yet static approach based on the 15-minute findings presented earlier. It is a different proposition than algorithms based on the total sentiment and volume of messages in the network[63]. While these algorithms base on mass and anonymous sentiments, the method presented in this work tries to exploit specific messages with high quality content from trusted sources. Intentionally, informative messages drive the public opinion and therefore this approach gains a speed-advantage over algorithms based on sentiments. Nonetheless, the proposed algorithm contains several weaknesses. *First*, the static purchase and sale timings are based on the minimal and maximal cumulative averages of all messages. As the findings are based on 1-minute intervals, the timings require optimization. The implementation of technical indicators (e.g. On-Balance Volume, Aroon Indicator and Stochastic Oscillator) can improve the return but require deeper analysis. *Second*, this research only investigated the 15-minute window. Therefore, it might be more profitable to hold the cryptocurrency for a longer duration. *Third,* the rate limit on the Twitter API to 900 requests per 15 minutes increase the likelihood to miss or react late to newly published messages. However, the presented version of the trading algorithm is slightly beneficial but holds too high risks for a practical use. The average standard deviation for the profitable clusters is 4% while the medians are zero. Nonetheless, the findings reveal high potential for future research. Especially a different time window can increase the profitability of algorithmic trading.

---

[63] *Colianni* et al. (2015)

In conclusion, the outlined results have shown a significant impact of informative Twitter messages from qualitative high sources within the first three minutes after the publication. Furthermore, sentiments and scope factors were found to have no influence on price movements. Lastly, this paper has provided a basic trading algorithm that was not yet beneficial but holds potential for further investigation and improvement.

# 7 Literaturverzeichnis

# Appendix A

| Factor | Short Description | Amount of variables | Variable type |
|---|---|---|---|
| **Sentiment** | Sentiment from the whole Twitter message calculated via the sentiment_by function of the sentimentr package in R | 5 evenly sized groups | Numeric |
| **Screen_Name** | The Twitter-ID | 108 | Text |
| **Tweet_Word** | A word that is in the whole Twitter message | 2744 | Text |
| **followers count** | Amount of followers, the Twitter account has | 5 evenly sized groups | Numeric |
| **friends count** | The number of users this account is following | 5 evenly sized groups | Numeric |
| **statuses count** | number of Tweets the account has posted | 5 evenly sized groups | Numeric |
| **favorites count** | The number of Tweets this user has liked | 5 evenly sized groups | Numeric |
| **retweet followers count** | Amount of followers, the retweeted account has | 5 evenly sized groups | Numeric |
| **retweet friends count** | Amount of people, the retweet account is following | 5 evenly sized groups | Numeric |
| **hashtag** | Hashtags used in the Twitter message | 227 | Text |
| **media_type** | Photos or Videos that are used in a Twitter message | 2 | Text |
| **mentions_screen_name** | Twitter-ID from mentions in a Twitter message | 206 | Text |
| **language** | Language of the Twitter message | 42 | Text |
| **quoted_text** | The whole text that was quoted from another Twitter User | 172 | Text |
| **quoted_screen-name** | The Twitter-ID from the quoted account | 29 | Text |
| **quoted followers count** | Amount of followers, the quoted account has | 5 evenly sized groups | Numeric |
| **quoted friends count** | Amount of people, the quoted account is following | 5 evenly sized groups | Numeric |
| **retweet_text** | The whole text that was retweeted from another Twitter User | 1021 | Text |
| **retweet screen name** | The Twitter-ID from the retweeted account | 206 | Text |